\documentclass[prl,groupedaddress,showpacs,nofootinbib,twocolumn]{revtex4}
\usepackage{graphicx}
\usepackage{amsmath}
\DeclareGraphicsExtensions{.pdf,.eps,.png,.jpg,.mps}
\usepackage{epstopdf}
\usepackage{amssymb}
\usepackage{color}
\newcommand{\ped}[1]{\ensuremath{_{\rm #1}}}
\newcommand{\apex}[1]{\ensuremath{^{\rm #1}}}

\begin{document}

\title{Weak Localization in Electric-Double-Layer Gated Few-layer Graphene}
\author{R. S. Gonnelli$^1$}\email{E-mail: renato.gonnelli@polito.it}
\author{E. Piatti$^1$}
\author{A. Sola$^{1,+}$}
\author{M. Tortello$^1$}
\author{F. Dolcini$^{1,2}$}
\author{S. Galasso$^1$}
\author{J. R. Nair$^1$}
\author{C. Gerbaldi$^1$}
\author{E. Cappelluti$^3$}
\author{M. Bruna$^4$}
\author{A. C. Ferrari$^4$}
\affiliation {$^1$Dipartimento di Scienza Applicata e Tecnologia, Politecnico di
Torino, 10129 Torino, Italy}
\affiliation {$^2$Istituto CNR-SPIN, Monte S.Angelo - via Cinthia, I-80126 Napoli, Italy}
\affiliation {$^3$Istituto dei Sistemi Complessi del CNR, 00185 Roma, Italy}
\affiliation {$^4$Cambridge Graphene Centre, University of Cambridge, Cambridge, CB3 OFA UK}
\pacs{72.80.Vp, 73.20.Fz, 73.22.Pr, 73.30.+y}
\begin{abstract}
We induce surface carrier densities up to $\sim7\cdot 10^{14}$cm$^{-2}$ in few-layer graphene devices by electric double layer gating with a polymeric electrolyte. In 3-, 4- and 5-layer graphene below 20-30K we observe a logarithmic upturn of resistance that we attribute to weak localization in the diffusive regime. By studying this effect as a function of carrier density and with ab-initio calculations we derive the dependence of transport, intervalley and phase coherence scattering lifetimes on total carrier density. We find that electron-electron scattering in the Nyquist regime is the main source of dephasing at temperatures lower than 30K in the $\sim10^{13}$cm$^{-2}$ to $\sim7 \cdot 10^{14}$cm$^{-2}$ range of carrier densities. With the increase of gate voltage, transport elastic scattering is dominated by the competing effects due to the increase in both carrier density and charged scattering centers at the surface. We also tune our devices into a crossover regime between weak and strong localization, indicating that simultaneous tunability of both carrier and defect density at the surface of electric double layer gated materials is possible.
\end{abstract}
\maketitle
\section{Introduction}
{\let\thefootnote\relax\footnotetext{\textsuperscript{+}present address: Istituto Nazionale di Ricerca Metrologica (INRIM), 10135 Torino, Italy}}
Electrolytic gating, initially developed for polymeric transistors (see Ref.\cite{Panzer05} and references therein), is now used to study the transport properties of a wide range of materials, from semiconductors\cite{Bayer05}, to insulators\cite{Yuan09} and superconductors\cite{Ueno08,Ye09,Ye12}. This technique induces orders-of-magnitude enhancement in the electric field at the sample surface, when compared to conventional solid gate techniques\cite{UenoReview2014,FujimotoReview2013,GoldmanReview2014,SaitoReview2016}. When a potential is applied between the sample and a counter electrode, the ions inside the polymeric (or liquid) electrolyte migrate and accumulate at the two surfaces, building up the so-called electric double layer (EDL)\cite{UenoReview2014,FujimotoReview2013}. This layer acts as a nanoscale capacitor, and allows one to obtain surface variations of the carrier density $n\ped {2d}$ up to$\sim10\apex{14}-10\apex{15}$cm$\apex{-2}$\cite{Ye09}, depending on the density of states of the sample and the electrochemical stability window of the electrolyte itself.

We firstly exploited an EDL to gate single layer graphene (SLG) in Ref.\cite{DasNatureNano2008} and bilayer graphene (BLG) in Ref.\cite{DasPRB09}. Since then, this technique has been widely applied to study the transport properties of SLG\cite{ChenJACS2009,Pachoud10,Efetov10} and few-layer graphene (FLG)\cite{Lui11,YePNAS,ChenNanoLett2012}. Refs.\cite{Pachoud10,Efetov10} studied the dominant scattering mechanisms in SLG for $n\ped {2d}$ up to 6 and 11$\cdot 10^{13}$cm$\apex{-2}$, respectively. Ref.\cite{Efetov10} reported a gate-tunable crossover of the resistivity from $\rho\propto \mathrm{T}$ to $\rho \propto \mathrm{T}^4$, in the range from 1.5K up to room temperature (RT). This constitutes a clear example of Bloch-Gruneisen behavior in two dimensions (2d) due to a crossover from small-angle to large-angle electron-phonon scattering\cite{HwangPRB2008}. Ref.\cite{Lui11} reported an electric field-induced, gate-tunable band gap in rhombohedral stacked three-layer graphene (3LG) for $n\ped {2d}\lesssim 1 \cdot 10^{13}$cm$\apex{-2}$. Ref.\cite{YePNAS} showed that the $n\ped {2d}$ achievable by electrolyte gating is large enough to populate the higher-energy bands in 2LG and 3LG, away from the charge neutrality point. Ref.\cite{ChenNanoLett2012} demonstrated a gate-tunable crossover from weak localization (WL) to weak anti-localization (WAL) in the low-T transport properties of 3LG on silicon carbide, without specifying the range of carrier densities achieved in their experiments.

We have been able to reach $n\ped{2d}$ up to $\sim3.5-4.5 \cdot 10^{15}$cm$\apex{-2}$ in thin films of Au and other noble metals with EDL gating by using a polymer electrolyte system (PES) of improved capacitance\cite{Daghero12,Tortello13} compared to previous reports\cite{UenoReview2014,FujimotoReview2013,GoldmanReview2014,SaitoReview2016}. By using the same PES, we were able to accumulate $n\ped{2d}>6 \cdot 10^{14}$cm$\apex{-2}$ in FLG flakes mechanically exfoliated from highly oriented pyrolytic graphite (HOPG)\cite{Gonnelli15}. One of the challenges at such high $n\ped{2d}$ is the observation of quantum phenomena, such as superconductivity. The induction of superconducting order in graphene would greatly benefit the development of novel device concepts\cite{Profeta12}, such as atomic-scale superconducting transistors\cite{DelahayeScience2003}, and superconductor-quantum dot devices\cite{DeFranceschiNatureNano2010,HuefnerPRB2009}. Superconductivity was reported in EDL gated MoS$\ped{2}$\cite{Ye12} and WS$\ped{2}$\cite{Jo15}. Refs.\cite{Profeta12,Margine14} theoretically predicted that this should also be the case for graphene. However, the experimental search for superconductivity in graphene via EDL-gating has so far been unsuccessful\cite{Efetov10,Gonnelli15}, while low-T superconductivity has been demonstrated by alkali-metal decoration\cite{LudbrookPNAS2015} and intercalation\cite{IchinokuraACSNano2016}.
\begin{figure*}
\centerline{\includegraphics[width=160mm]{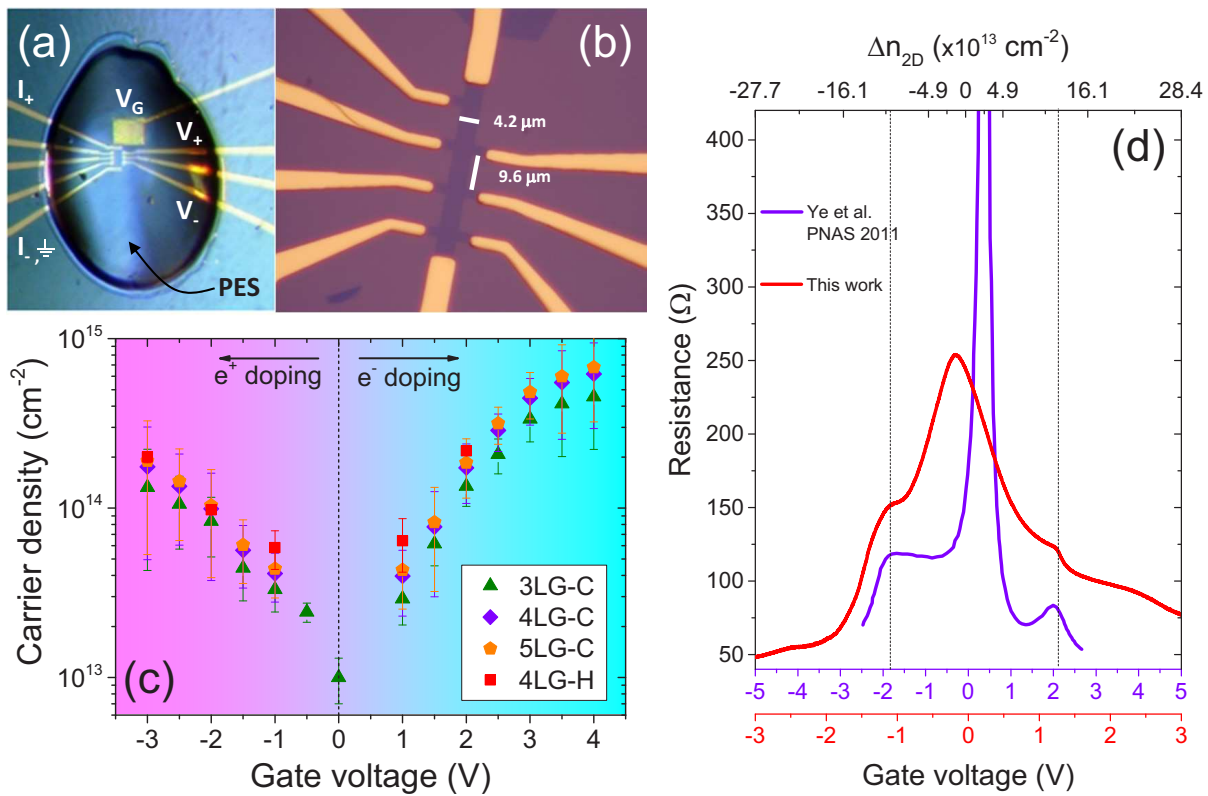}}
\caption{a) FLG device covered by drop-cast PES; b) Zoom in the region of the Hall-bar shaped FLG channel showing current and voltage contacts; c) n$\ped{2d}$ as a function of V$\ped{G}$. In the legend, C stands for "determined by double-step chronocoulometry", while H means "determined by Hall-effect measurements"; d) Resistance curve as a function of V$\ped{G}$ and of the induced carrier density $\Delta n\ped{2d} = n\ped{2d}(\mathrm{V}\ped{G}) - n\ped{2d}(0)$ of one of our 3LG devices compared with Ref.\cite{YePNAS}.}
\label{fig:1}
\end{figure*}

Here, we focus on the transport properties of EDL-gated 3LG, 4LG and 5LG in the 4-30K range. We find a small ($\lesssim 1.5\%$), gate tunable, upturn in the sheet resistance, $R_S$, below $\sim$20-30K. This shows a logarithmic T dependence, and its intensity is inversely correlated with the carrier density. We demonstrate by magnetoresistance measurements that WL in the diffusive regime is at the origin of the upturn. We find that the experimental behavior of FLGs agrees well with the theoretical models developed for SLG and 2LG. Combining experiments with \textit{ab-initio} calculations we extract the dependence of the characteristic scattering lifetimes with both T and carrier density. From the dephasing lifetime dependence on T, we find that the dominant inelastic scattering mechanism for $\mathrm{T}\lesssim$20K is the electron-electron scattering with small momentum transfer (Nyquist term)\cite{Altshuler_elscat}. From its dependence on carrier density, we show how EDL gating brings the system in a crossover condition from WL to strong localization, and how this is gate-tunable in the case of 5LG. Our results demonstrate quantum coherent transport in 4LG and 5LG for carrier densities in excess of$\sim10^{13}-10^{14}$cm$\apex{-2}$.
\section{Fabrication and characterization}\label{sect:fabr_char}
\begin{figure*}
\centerline{\includegraphics[width=170mm]{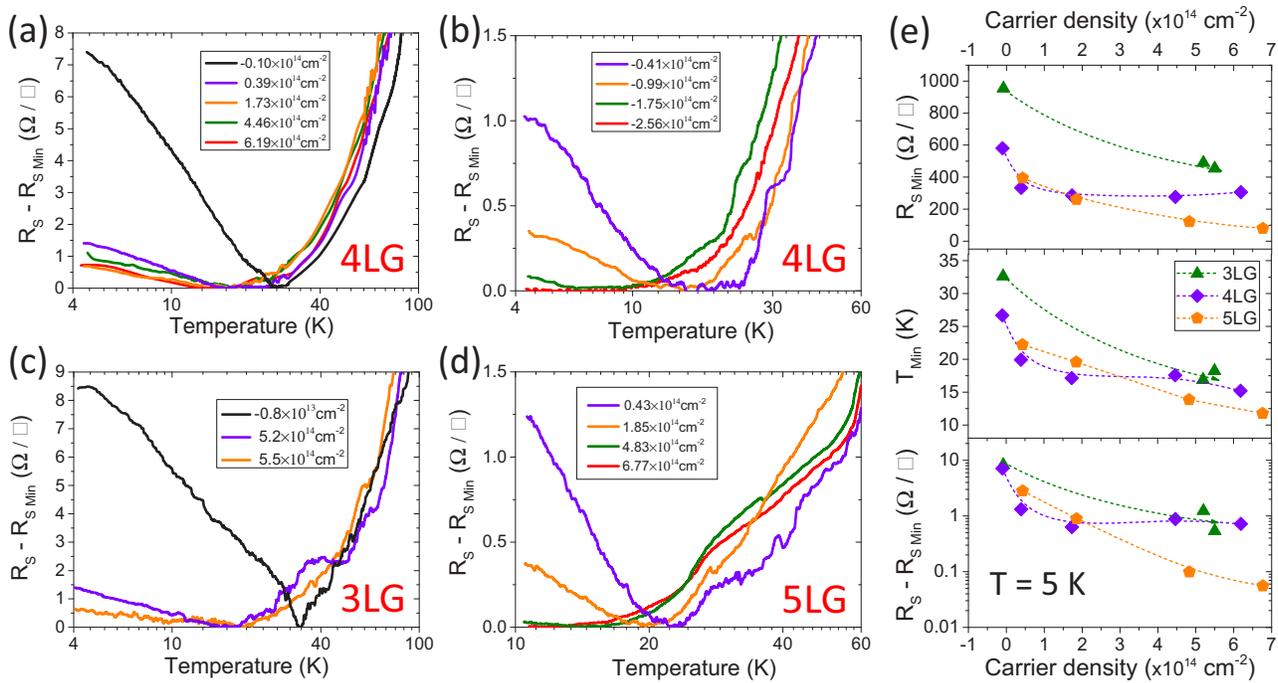}}
\caption{Semi-logarithmic plot of the increase of R$\ped{S}$ with respect to its minimum R$\ped{S Min}$ as a function of T in a) 4LG for different values of electron doping; b) the same device for different values of hole doping; c) 3LG and d) 5LG for different values of electron doping. e) Carrier density dependencies of R$\ped{S Min}$ (top panel), temperature T$\ped{Min}$ where the minimum is reached (central panel), and R$\ped{S}$ increase at $\mathrm{T} = 5$K (bottom panel), for different N. Dashed lines are guides to the eye.} \label{fig:2}
\end{figure*}
FLG flakes are prepared by micromechanical exfoliation of graphite on Si/SiO$_2$\cite{NovoselovPNAS2005,BonaccorsoMatToday2012}. The number of layers, N, is determined by a combination of Raman spectroscopy\cite{Ferrari2006,Ferrari_NNano2013} and optical microscopy\cite{CasiraghiNanoLett2007}. All FLGs are Bernal stacked. EDL field-effect devices are then prepared by electron beam lithography and deposition of Cr/Au contacts in the Hall bar configuration, as shown in Fig.\ref{fig:1}b. A coplanar gate contact is deposited on the side. A protective poly(methyl methacrylate) (PMMA) layer is spin-coated, patterned and hard-baked on the completed devices in order to avoid interactions between the electrolyte and the metallic leads. A reactive mixture precursor of the PES is then drop-cast onto the whole device and UV-cured, to get a cross-linked photopolymer electrolyte\cite{Daghero12,Gonnelli15}, Fig.\ref{fig:1}a. The precursor viscous liquid reactive mixture is composed of methacrylic oligomers and lithium salt: bisphenol A ethoxylate dimethacrylate (BEMA; average Mn: 1700, Aldrich), poly(ethylene glycol)methyl ether methacrylate (PEGMA; average Mn: 475, Aldrich), and 10\% of lithium bis(trifluoromethanesulfonyl)imide (LiTFSI) along with the addition of 3 wt\% 2-hydroxy-2-methyl-1-phenyl-1-propanon as the free radical photoinitiator (Darocur 1173, Ciba Specialty Chemicals). No annealing is performed in the time between contact deposition and PES casting, and the devices are stored in a desiccator under low vacuum ($\sim10^{-1}-1$mbar) to prevent contamination. Four-wire resistance measurements are done under high vacuum ($<10^{-5}$mbar) in the chamber of a Cryomech pulse-tube cryocooler via a Keithley 6221/2182A current source/nanovoltmeter assembly. The gate potential is controlled through a Keithley 2410 sourcemeter, whose negative electrode is set at the current drain contact. Gate leakage is monitored to ensure that no electrochemical reactions occur at the graphene/EDL interface. Magnetoresistance measurements are performed in an Oxford Instruments helium cryostat equipped with a superconducting magnet.

We estimate $n\ped{2d}$ as a function of the gate voltage, V$\ped{G}$, from the measured gate current through double-step chronocoulometry\cite{ScholzInzelt} (DSCC), as shown in Fig.\ref{fig:1}c for 3LG, 4LG and 5LG (green triangles, violet diamonds and orange pentagons, respectively) and described in detail in Ref.\cite{Gonnelli15}. This technique is based on the fact that, in absence of chemical reactions at the electrodes, the current flowing through the electrolyte can be split into two contributions\cite{ScholzInzelt}. The first, due to the EDL build-up, decays exponentially over time\cite{ScholzInzelt}. The second, due to ion diffusion through the bulk of the electrolyte\cite{ScholzInzelt}, decays as the square root of time\cite{ScholzInzelt}. By fitting the gate current response to a step-like application and removal of V$\ped{G}$ with a function that separately accounts for both effects, the two contributions can be distinguished and, thus, the EDL charge estimated. Further details on the technique can be found in Refs.\cite{Daghero12,Gonnelli15,ScholzInzelt}. These results are checked with independent Hall-effect measurements performed on a 4LG device (red squares in Fig.\ref{fig:1}c). We conservatively use the DSCC values in the full V$\ped{G}$ range.

We also compare the RT resistance as a function of V$\ped{G}$ in a 3LG with the earlier measurements of Ref.\cite{YePNAS}. Fig.\ref{fig:1}d shows how, in both cases, regions of sharp non-monotonic behavior are present in the resistance curves away from the Dirac point. The position of these upturns is associated with a specific charge density, such that the Fermi level, E$\ped{F}$ is crossing the bottom of the split-off bands\cite{YePNAS}, and this fact can be used for voltage-carrier density calibration. This comparison indicates that our PES allows us to reach the split-off bands at lower potential than the ionic liquid of Ref.\cite{YePNAS}, thus supporting a higher charge induction capability.
\section{Results and Discussion}\label{sect:RT}
We first measure the four-contact resistance in the 4-290K range as a function of N and $n\ped{2d}$. In each measurement, V$\ped{G}$ is applied at RT and the sample is then cooled to a base T$\sim$4-10K, depending on sample. Any further modification of $n\ped{2d}$ below the PES glass transition at$\sim$230K\cite{Gonnelli15} is prevented by the quenched ion motion\cite{Ueno08,Ye12}. The samples are then allowed to spontaneously warm up, taking advantage of the absence of T fluctuations associated with the cryocooler's thermal cycles. We reported a detailed analysis of the T dependence of R$\ped{S}$ for these device in the 30-290K range in Ref.\cite{Gonnelli15}. Here we focus on the anomalous behavior that emerges in the 4-30K range.

For all the three N, a logarithmic upturn is observed for T$\lesssim 30$K in the resistance curves for samples in the metallic regime, as shown in Fig.\ref{fig:2}. Figs.\ref{fig:2}a-d plot the low-T ($\lesssim 100$ K) variations of R$\ped{S}$ with respect to its minimum value R$\ped{S Min}$ for 3LG, 4LG and 5LG as function of T in a semi-logarithmic scale. R$\ped{S Min}$ decreases with increasing n$\ped{2d}$ for all N, and saturates for 4LG (see top panel of Fig.\ref{fig:2}e). Both the amplitude of the logarithmic upturn (R$\ped{S}(5 \mathrm{K})-$R$\ped{S Min}$) and the T$\ped{Min}$ at which R$\ped{S Min}$ is reached strongly depend on $n\ped{2d}$ and N, as shown in the central and bottom panels of Fig.\ref{fig:2}e. Similar logarithmic upturns in the resistance of conductors in the diffusive regime, such as Si MOSFETs\cite{WheelerPRL1982}, GaAs heterostructures\cite{ChoiPRB1986,Hansen}, and transition metal alloys\cite{SarachikPR1964} are usually attributed to Kondo effect\cite{Kondo1964}, WL\cite{BergmannWL,LarkinWL}, and electron-electron (e-e) interactions\cite{Altshuler_elinter,Altshuler_elinter_Rev,FukuyamaEE}, or to a superposition of any of the previous sources\cite{BeenakkerWL}.

\begin{figure}
\centerline{\includegraphics[width=80mm]{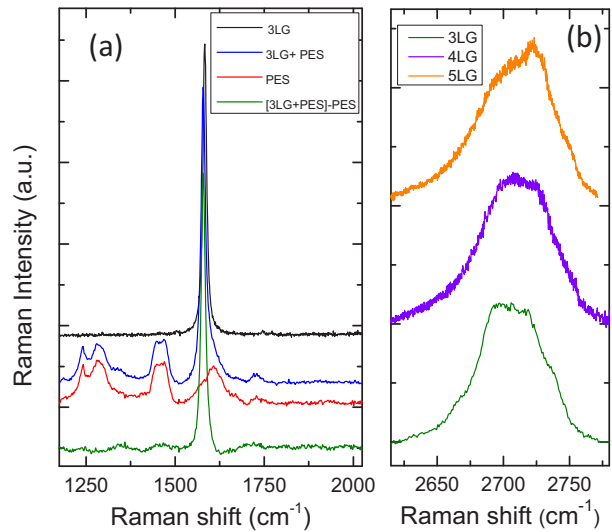}}
\caption{a) Raman spectrum measured at $514$nm for a 3LG device, before (black) and after (blue) PES drop-casting, together with the Raman response from the PES (red) and the spectrum with the subtracted background (green). b) Raman 2D region for 3LG (green), 4LG (violet) and 5LG (orange).}
\label{fig:3}
\end{figure}

We first consider the Kondo effect, which is caused by carrier scattering from diluted magnetic impurities\cite{SarachikPR1964,Kondo1964}. Unlike Ref.\cite{ChenKondo}, we do not intentionally introduce any defect in our samples. We thus employ Raman spectroscopy to monitor the amount of Raman active defects in our devices, before and after PES drop-casting. Raman measurements are carried out at RT in a Renishaw InVia microspectrometer equipped with a 100X objective. The spot size is$\sim1\mu$m, the excitation wavelength is 514nm and the incident power is kept well below 1mW in order to avoid heating. Fig.\ref{fig:3}a shows the Raman spectrum of a 3LG device. The PES shows several peaks between$\sim$1000 and 1500cm$\apex{-1}$. The broad peak between$\sim$1400 and 1500cm$\apex{-1}$ can be assigned to the methylene bending mode\cite{ReyElecActa1998}, the peak at$\sim$1280cm$\apex{-1}$ to methylene twisting\cite{ChrissopoulouMacmol2011}, and that at$\sim$1243cm$\apex{-1}$ to either trifluoromethyl stretching\cite{ReyElecActa1998} or methylene twisting\cite{ChrissopoulouMacmol2011}.

We perform a background subtraction by normalizing to the low-frequency PES peaks near the D peak. This gives an upper limit for the ratio between the intensity of the D peak and the intensity of the G peak, I(D)/I(G),$\sim$0.015, which leads to an estimated maximum defect density$\sim3\cdot 10^9$cm$\apex{-2}$\cite{CancanoNanoLett2011,BrunaACSNano2014}. For comparison, in order to obtain a sizeable Kondo effect ($\delta \rho / \rho \simeq 6\%$ at T=0.3K), in Ref.\cite{ChenKondo} it was necessary to induce a defect density$\sim3\cdot 10^{11}$cm$\apex{-2}$, which produced a corresponding I(D)/I(G)$\sim$3.8\cite{ChenKondo}. Our values indicate that our devices are at least two orders of magnitude less defective, and we thus discard the possibility of a contribution from Kondo effect to the resistance upturn.
\subsection{Weak Localization in Few-layer Graphene}\label{sect:WL}
WL arises from the constructive interference between pairs of time-reversed trajectories of electrons elastically scattering in a closed loop\cite{BeenakkerWL}. This increases the probability of charge-carrier backscattering in a conductor in the diffusive regime\cite{BergmannWL,LarkinWL}, thus reducing the conductance\cite{BeenakkerWL}. In order for WL to appear, phase coherence must be maintained throughout the entire closed loop followed by the electrons undergoing the elastic scattering that is described by a characteristic lifetime $\tau\ped{e}$\cite{BergmannWL}. WL is thus suppressed by an increased probability of inelastic scattering associated with the T increase\cite{BergmannWL}. The average time interval in which phase coherence is maintained is called phase coherence lifetime $\tau\ped{\varphi}$\cite{BeenakkerWL}. Typically, two inelastic processes determine $\tau\ped{\varphi}$\cite{BeenakkerWL}. Electron-phonon (e-ph) scattering\cite{Taboryski}, which is important for higher T\cite{BeenakkerWL}, and electron-electron (e-e) scattering\cite{Narozhny,Altshuler_elscat}, which dominates for lower T\cite{BeenakkerWL}. The T that marks a crossover between these two scattering processes depends on the transport properties of each material\cite{BeenakkerWL}, and it is typically of the order of few tens K\cite{BeenakkerWL}.

In a diffusive system a conductivity suppression at low T may also be caused by e-e interactions, as a result of the diffraction of an electron wave by the oscillations in the electrostatic potential generated by the other electrons\cite{BeenakkerWL,Altshuler_elinter,Altshuler_elinter_Rev,FukuyamaEE}. A widely exploited approach to discriminate between these two contributions is the measurement of the sample magnetoconductance\cite{BeenakkerWL,Altshuler_elinter,Altshuler_magnetoR}. In the presence of a perpendicular magnetic field B, time-reversal symmetry is broken\cite{BeenakkerWL}. A phase difference arises between any two time-reversed electron trajectories, whose constructive interference, in principle able to localize electrons\cite{BeenakkerWL}, is therefore broken\cite{Altshuler_magnetoR}. As a consequence\cite{BeenakkerWL}: i) a negative magnetoconductance is associated with WAL\cite{Hikami_WLB}; ii) a vanishing magnetoconductance is associated with e-e interaction (insensitive to B)\cite{Altshuler_magnetoR}; iii) a positive magnetoconductance is associated with WL\cite{Altshuler_magnetoR,Hikami_WLB}.
\begin{figure}
\centerline{\includegraphics[width=80mm]{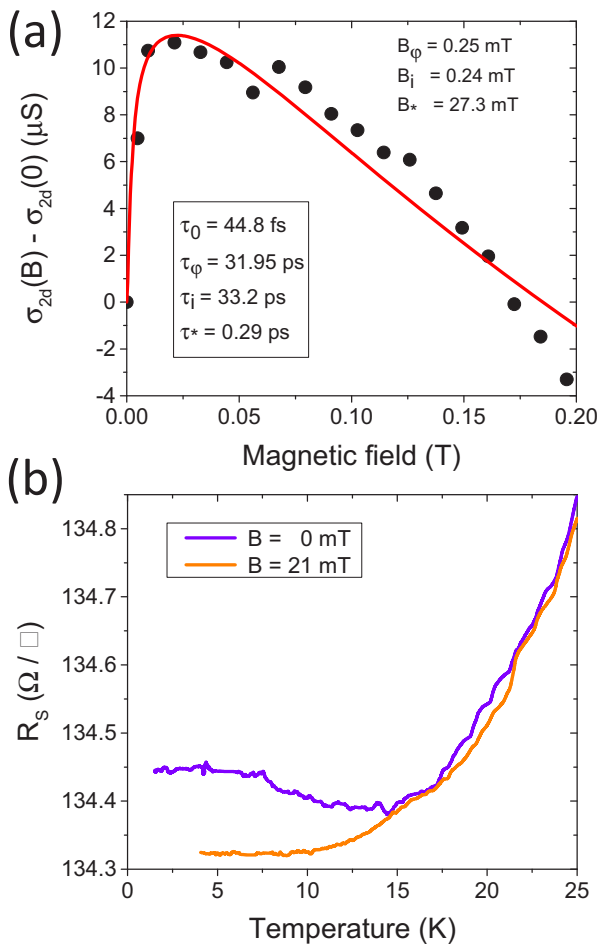}}
\caption{a) B dependence of conductance relative to the value at zero field (black circles) and fit by using Eq.\ref{WL-magfield} (red solid line) for 3LG. b) R$\ped{S}$ as a function T for the 3LG device of Fig.\ref{fig:4}a for B=0.021T and 0.} \label{fig:4}
\end{figure}

In order to determine the origin of the resistance upturn we thus measure the magnetoconductance of a 3LG device at 2K. We find it positive, and reaching a maximum at $B_\varphi=2\hbar/4ev\ped{F}\apex{2}\tau\ped{e}\tau\ped{\varphi}=0.021$T, as shown in Fig.\ref{fig:4}a, where the magnetoconductance relative to the value at zero field is plotted as a function of B. In order to rule out even a partial contribution from e-e interactions, we also measure the device resistance while heating the sample under $\mathrm{B}=\mathrm{B}_\varphi$. Fig.\ref{fig:4}b shows that the logarithmic upturn is suppressed within the noise level. This behavior is reversible upon removal of B. The logarithmic upturn saturates for $\mathrm{T}\lesssim$4K, a feature usually assigned to the effect of finite channel size\cite{Tikhonenko09}. These data support the conclusion that, in our devices, WL is the source of the measured resistance upturn and that the logarithmic correction to the conductivity due to e-e interaction can be disregarded. This is in contrast with earlier reports on SLG, where both contributions were shown to be relevant\cite{KozikovPRB2010}.

The carrier lifetimes determine the WL behavior\cite{BeenakkerWL,Hikami_WLB}, and can thus be obtained from Fig.\ref{fig:4}a\cite{McCann06,Falko07}. In SLG it was predicted that, due to lattice symmetries, an additional $\pi$ Berry phase is accumulated by charge carriers scattering in closed loops\cite{Zhang05,Suzuura02,Suzuura03}. In 2LG such additional Berry phase is instead $2\pi$\cite{Novoselov06,Falko07}. This means that in SLG the quantum interference due to localization is always destructive\cite{Ando98,Ando2005}, leading to an anti-localization behavior\cite{Hikami_WLB}, while in 2LG the effect of the additional phase cancels out, and the material should retain the standard WL behavior\cite{Falko07}. In addition, WAL, where the underlying elastic scattering mechanism is due to charged impurities, is strongly suppressed by intervalley scattering and trigonal warping\cite{McCann06}. Furthermore, Refs.\cite{McCann06,Falko07} showed that SLG and 2LG have a further quantum correction to the Drude resistivity, always localized, due to elastic intervalley scattering, and not affected by trigonal warping. As a consequence, various localization behaviors were reported\cite{Tikhonenko09,Tikhonenko08,Hikami_WLB,Wu07,Berger06,Morozov06,Ki08,Liao10}, depending on the sample preparation and on the explored regime.

To the best of our knowledge, a WL model for $\mathrm{N}>2$ has not yet been developed. Earlier WL reports in FLG made use of models developed for SLG and 2LG\cite{LiuPCCP2011,WangCarbon2012,ChuangCurrApplPhys2014}. Furthermore, the simultaneous presence of both Dirac-like and parabolic bands in odd-N flakes\cite{Nakamura08,Koshino10} makes an a-priori determination of the total Berry phase difficult. Here we show that the correction to the conductance due to the Berry phase is suppressed in all our experiments. This allows us to conclude that only elastic intervalley scattering is relevant for quantum interference.

Following the theoretical approach of Ref.\cite{Falko07}, the T dependence of the quantum correction to the 2d conductance is determined by the T behavior of the phase coherence lifetime $\tau_\varphi$, as:
\begin{equation}
\delta\sigma\ped{WL}(\mathrm{T})= -\frac{e^2}{\pi h}\ln\left[1+2\frac{\tau_\varphi(\mathrm{T})}{\tau_i}\right] - \delta_0(\tau_\varphi,\tau\ped{tr},\tau_*)
\label{WL-temp}
\end{equation}
where $\tau_i$ is the intervalley scattering lifetime, $\tau\ped{tr}=2\tau_e$ is the transport lifetime (associated with scattering from charged impurities\cite{Tikhonenko09,Tikhonenko08}) and $\tau_*$ is an effective lifetime associated with intravalley scattering and trigonal warping. On the right hand side of Eq.\ref{WL-temp}, the first term arises from intervalley scattering, and resembles the conventional weak localization correction\cite{BeenakkerWL}, apart from the factor two and the different elastic lifetime involved. Since the argument of the logarithm is always greater than one, this term is always negative. The second term, arising from intravalley scattering, is positive or negative depending on the Berry phase, and is strongly suppressed the larger the value of $\tau_*^{-1}$\cite{McCann06,Falko07}. The B dependence of the quantum correction to the conductance can be written as\cite{McCann06,Falko07}:
\begin{equation}
\begin{split}
\delta\sigma\ped{WL}(\mathrm{B}) =& \frac{e^2}{\pi h} \left[F\left(\frac{\mathrm{B}}{\mathrm{B}_\varphi}\right)-F\left(\frac{\mathrm{B}}{\mathrm{B}_\varphi+2B_i}\right)\right]+ \\
\\& +\delta(\mathrm{B},\mathrm{B}_\varphi,B_*)
\end{split}
\label{WL-magfield}
\end{equation}
where:
\begin{equation}
F(z) = \ln(z) + \psi(1/2 + 1/z)\nonumber
\label{F}
\end{equation}
and
\begin{equation}
B\ped{\varphi,i,*} = (\hbar/4D\ped{tr}e)\tau\ped{\varphi,i,*}^{-1}\nonumber
\label{B}
\end{equation}
$\psi$ being the digamma function\cite{Falko07}, $D\ped{tr}=v\ped{F}^{2}\tau\ped{tr}$ the diffusion coefficient\cite{Falko07}, and $v\ped{F}$ the Fermi velocity. We obtain $\tau\ped{tr}$ by combining the measurement of $R\ped{SMin}$ with \textit{ab-initio} DFT calculations (see Eq.\ref{tau_transport}). This gives the values $\tau\ped{tr}=44.8$fs and $D\ped{tr}=448$cm$\apex{2}$s$\apex{-1}$ that will be used in later calculations.

For the data in Fig.\ref{fig:4}a (3LG), the Berry phase is $\pi$ and $\delta(\mathrm{B},\mathrm{B}_\varphi,\mathrm{B}_*)=-(4e^2/\pi h)F(\mathrm{B}/(\mathrm{B}_\varphi+\mathrm{B}_*))$. We therefore use Eq.\ref{WL-magfield} to fit $\sigma\ped{2d}$ as a function of B in the 0-0.2T range (red solid curve in Fig.\ref{fig:4}a). This gives $\tau_\varphi=31.95$ps, $\tau_i=33.2$ps and $\tau_*=0.29$ps. Thus, we are in the condition $\tau_*\ll\tau\ped{\varphi,i}$, where the zero-field antilocalization term $\delta_0$ is suppressed to the point of being negligible\cite{McCann06,Falko07}.

We can compare these results with the existing literature at T=4K, immediately before the signal saturation for T$\lesssim$4K, shown in Fig.\ref{fig:4}b. Ref.\cite{Tikhonenko08} presents a WL study in SLG for different carrier densities, expressed in terms of the characteristic lengthscales $L\ped{\varphi,i,*} = (D\ped{tr}\tau\ped{\varphi,i,*})^{1/2}$. For $n\ped{2d}\simeq 1.5\cdot 10^{12}$cm$\apex{-2}$ and T=4K, they report $L_\varphi=1.5\mu$m, $L_i=0.8\mu$m and $L_*=75$nm. This is the value of $n\ped{2d}$ closest to that in our devices, even if still at least one order of magnitude smaller. The characteristic lifetimes in our device give $L_\varphi=0.93\mu$m, $L_i=0.95\mu$m and $L_*=88$nm, in agreement with the existing data\cite{Tikhonenko08}.

Since the T dependence of $\tau_\varphi$ contains information on the most relevant inelastic scattering mechanisms\cite{BeenakkerWL,Hikami_WLB} and following the previous results, we can use Eq.\ref{WL-temp} with $\delta_0=0$ to describe the low-T ($\lesssim$20-30K) part of the $R_S$ as a function of T curves in Fig.\ref{fig:2}. This approach is sound, since the smallest doping in our measurements is$\sim1 \cdot 10^{13}$cm$\apex{-2}$, a large value for solid-dielectric gating experiments\cite{UenoReview2014,FujimotoReview2013}, and we do not find anti-localized behavior at these doping levels. Refs.\cite{Tikhonenko09,Tikhonenko08,Baker_lphi} also reported that doping strongly enhances $\tau_*^{-1}$ (thus reducing $\delta_0$), since trigonal warping becomes more relevant the higher the energy separation from the Dirac point. This makes our approximation $\delta_0=0$ valid also for higher doping.

We can thus write the complete expression for the conductance of our FLG devices, valid for T$\lesssim 90$K:
\begin{equation}
\sigma\ped{2d}=\sigma\ped{Drude}+\delta\sigma\ped{WL}
\label{sigma_2d}
\end{equation}
where $\sigma\ped{Drude}$ is\cite{Cappelluti_sigma}:
\begin{equation}
\sigma\ped{Drude}=\frac{e^2}{4}\tau \sum_i (\nu\ped{k\ped{i}} \nu\ped{s} v\ped{F\ped{i}}\apex2 N\ped{i}) = e\apex2 \tau P(n\ped{2d})
\label{sigma_B}
\end{equation}
Here $\nu\ped{k\ped{i}}$ and $\nu\ped{s}$ are the valley and spin degeneracies, respectively, $v\ped{F\ped{i}}$ is the Fermi velocity and $N\ped{i}$ the DOS at the Fermi level of the i-th band. Taking into account the Matthiessen rule\cite{MatthiessenRule} on the carrier lifetime at finite T we have:
\begin{equation}
\tau^{-1} = \tau\ped{tr}^{-1} + \tau\ped{inel}^{-1}
\end{equation}
where $\tau\ped{inel}$ is the T-dependent inelastic scattering lifetime. Substituting in Eq. \ref{sigma_2d} we get:
\begin{equation}
\sigma\ped{2d}=e^2 P(n\ped{2d})\cdot(\tau\ped{tr}^{-1}+\tau\ped{inel}^{-1} )^{-1} - \frac{e^2}{\pi h} \ln\left(1+2 \frac{\tau_\varphi}{\tau_i}\right)
\label{full_cond}
\end{equation}
In order to proceed further, we need to determine the prefactor $P(n\ped{2d})$, the transport lifetime $\tau\ped{tr}$ and the T dependence of $\tau_\varphi$.

We estimate $P(n\ped{2d})$ by computing the FLG bandstructure through \textit{ab-initio} Density Functional Theory (DFT) for all our doping levels. We also perform a consistency check of the DFT estimations by running independent self-consistent tight-binding calculations for the doped 4LG. Here we present the results for Bernal stacked 4LG and 5LG. As a first approximation, we assume these systems to be isolated, i.e. we consider the effects of the substrate to be negligible.
\begin{figure}
\centerline{\includegraphics[width=90mm]{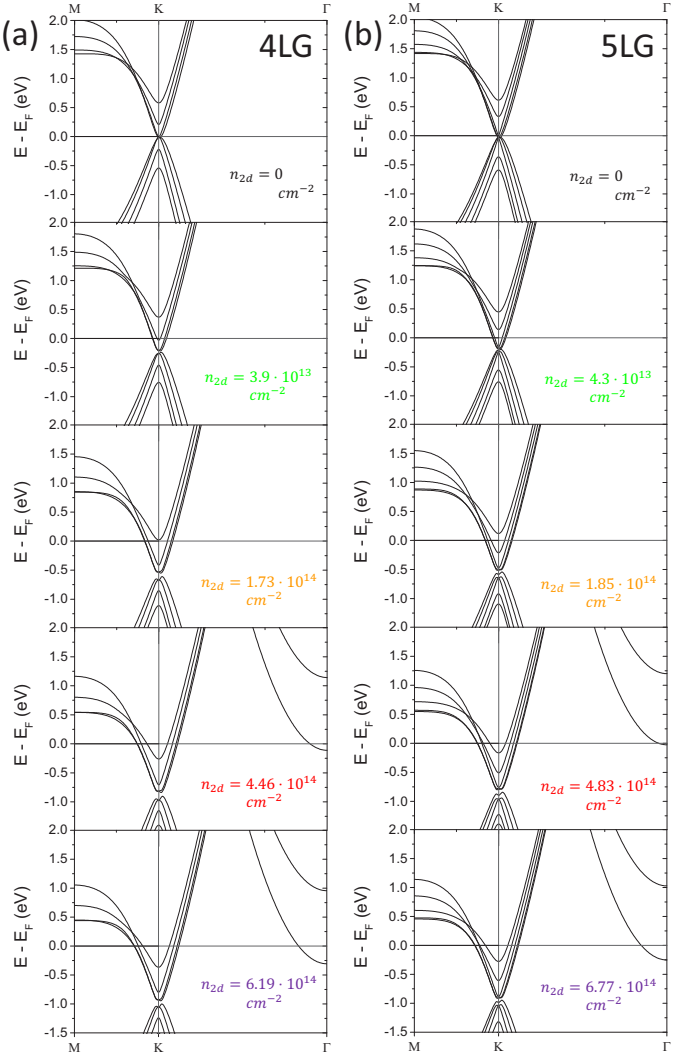}}
\caption{Band structure evolution with increase of doping in 4LG (a) and 5LG (b), for selected values of n$\ped{2d}$. For n$\ped{2d}\lesssim 2\cdot10\apex{14}$cm$\apex{-2}$, the effect of the extra carriers is comparable to a rigid-band shift. Larger doping induces new structures.}
\label{fig:bands}
\end{figure}

We first investigate the band-structures of FLG within the all-electron, full-potential, linear augmented plane wave (FP-LAPW) method as implemented in the ELK code\cite{ELK}. The local density approximation (LDA)\cite{Perdew81} is used to describe the exchange and correlation. We fix the in-plane lattice constant to a=2.46\AA, i.e. the experimental in-plane lattice constant of graphite\cite{SlonczewskiPR1958}. To model FLG we consider a three-dimensional supercell with a very high value of the c lattice constant so that the periodic images of the structures are at least 10\AA\ apart, in order to avoid interactions (e.g., for 4LG we take c=40a.u.) while the layer-layer distance is taken to be 3.35\AA\ independent of N. Doping is simulated by adding electrons to the systems, together with a compensating positive background, as in Refs.\cite{GeLiu,Margine14} (Jellium model). The Brillouin zone is sampled with a $28\times28\times1$ k-point mesh. The radius of the muffin-tin spheres for the carbon atoms are taken as 1.342$a_0$, where $a_0$ is the Bohr radius. We set R$\ped{MT}$K$\ped{max}$=8, where R$\ped{MT}$ is the smallest muffin-tin radius, K$\ped{max}$ is a cutoff wave vector and the charge density is Fourier expanded up to a maximum wave vector G$\ped{max} = 12 a_0$. The convergence of self-consistent field calculations is attained with a total energy tolerance of $10\apex{-8}$ Hartree.

Fig.\ref{fig:bands} plots the results for both 4LG (panel a) and 5LG (panel b) for different doping. The extra induced charge shifts E$\ped{F}$ away from the Dirac point, eventually crossing multiple bands at K and K'. In all cases, the crossing occurs well within regions where the energy dispersion is nearly linear. For the highest V$\ped{G}$ (+3 and +4V), however, an additional important feature appears in the bandstructure. The large induced charge density brings down a further band at $\Gamma$, which becomes populated and can thus contribute to the transport properties. The dispersion of this new band is fully parabolic, and its appearance is reminiscent of that of the interlayer band induced by alkali metal intercalation in graphite intercalated compounds (GICs)\cite{Profeta12}.

Our approach neglects the intense electric field at the EDL/graphene interface, that in transition-metal dichalcogenides increases with doping and affects the effective mass value\cite{Brumme15}. The simplified Jellium model was shown to capture the fundamental physics of doped SLG\cite{Margine14} and single-layer MoS$\ped{2}$\cite{GeLiu}. We thus perform a consistency check for 4LG by comparing DFT with tight-binding calculations where the screening of the electric field by the charge carriers is accounted for. We compare the same terms $\sum_i v\ped{F\ped{i}}^2 N_i$, where $i$ represent the band index, $v\ped{F\ped{i}}$ are the Fermi velocities and  $N_i$ are the density of the state at E$\ped{F}$. The $v\ped{F\ped{i}}$ and $N\ped{i}$ values obtained by tight-binding and DFT are within 30\%. This suggests that our DFT approach is correct at least to a first-order approximation.

These calculations allow us to obtain both the DOS and the electron velocity of the populated bands at E$\ped{F}$ which, in turn, according to Eq.\ref{sigma_B}, determine $P(n\ped{2d})$. Once $P(n\ped{2d})$ is known, $\tau\ped{tr}$ can be evaluated. Fig.\ref{fig:4}b shows that, in the absence of WL, the resistance saturates to a constant value for T$\lesssim$8K. Thus, the saturation Boltzmann conductance $e^2 \tau\ped{tr} P(n\ped{2d})$ can be estimated from the minimum R$\ped{S Min}$ before the onset of the logarithmic upturn. This underestimates the saturation conductance (and thus the scattering time) by less than 1\%, which is negligible compared to the other sources of uncertainty in the evaluation of $\tau\ped{tr}$. This approach allows us to compute $\tau\ped{tr}=2\tau\ped{e}$ as a function of n$\ped{2d}$ in all our devices as:
\begin{equation}
\tau\ped{tr}(n\ped{2d})=\frac{\mathrm{R}\ped{S Min}^{-1}(n\ped{2d})}{e^2 P(n\ped{2d})}
\label{tau_transport}
\end{equation}
\subsection{Scattering lifetimes vs. Temperature}
\begin{figure}
\centerline{\includegraphics[width=80mm]{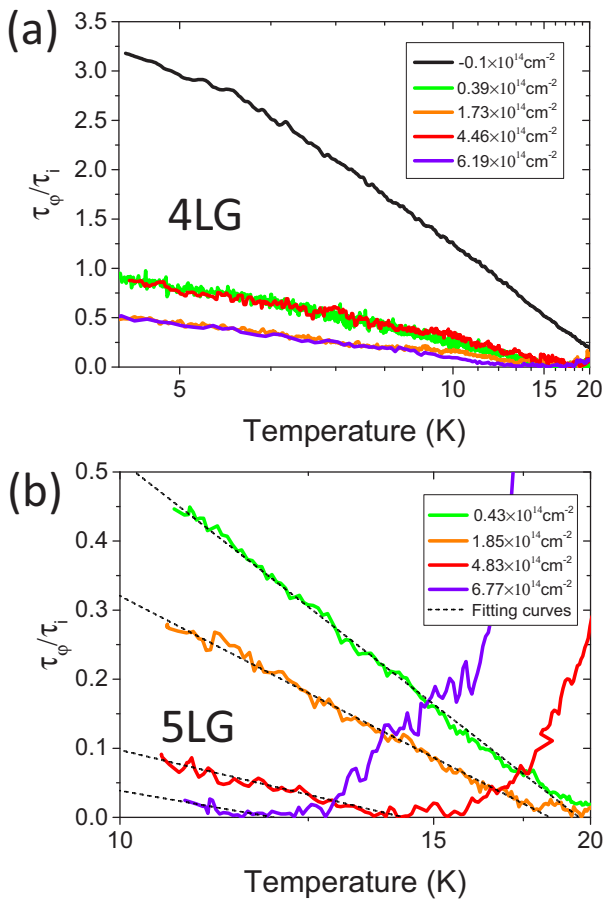}}
\caption{a) $\tau_\varphi/\tau_i$ as a function of 1/T obtained from the low-T data (4$<$T$<$20K) of Fig.\ref{fig:2}a (4LG); b) The same as in a) but for the low-T data of Fig.\ref{fig:2}d (5LG). Dashed lines are the best fits to the experimental curves according to the model described in the main text. Best fits for 4LG are omitted for clarity.}
\label{fig:5-1}
\end{figure}
We now consider the T dependence of $\tau_\varphi$. This can be done by using Eq.\ref{WL-temp} without the $\delta\ped{0}$ contribution (since $\tau_*\ll\tau\ped{\varphi,i}$) in the analysis of the low-T part (4$<$T$<$20K) of the curves of Fig.\ref{fig:2}. By inverting this simplified expression, we obtain the T dependence of $\tau_\varphi(\mathrm{T})/\tau\ped{i}$, as shown in Fig. \ref{fig:5-1}a (4LG) and b (5LG). Since $\tau\ped{i}$, for T$\lesssim$30K, can be assumed almost independent of T\cite{Tikhonenko08}, the T dependence of $\tau_\varphi/\tau_i$ is the same as that of $\tau_\varphi$. By expressing $\tau_\varphi\propto \mathrm{T}^{-\mathrm{p}}$, we associate the p value to specific scattering processes. For T lower than the Bloch-Gruneisen T\cite{HwangPRB2008}, e-ph scattering gives p=3 for a standard 2-dimensional electron gas (2DEG)\cite{Taboryski} and p=4 for SLG\cite{HwangPRB2008}. The e-e scattering gives p=1 for the Nyquist (small momentum exchange) process\cite{Altshuler_elscat} and p=2 for the Coulomb (large momentum exchange) process\cite{Narozhny}. As shown in Fig.\ref{fig:5-1}, the $\tau_\varphi/\tau\ped{i}$ curves show almost no superlinear behavior in the WL region. Nyquist e-e scattering is thus the dominant mechanism in this T range. This is consistent with the existing literature on SLG\cite{Tikhonenko09} and 2LG\cite{GorbachevPRL2007,ChenPhysicaB2011}. In Ref.\cite{Gonnelli15} we reported that the T dependence below 100K has a linear plus a quadratic contribution, the linear being dominant for T$\ped{Min} \lesssim T \lesssim$30K (see Figs.4a,5c of Ref.\cite{Gonnelli15}) and assigned this to competing Coulomb and Nyquist e-e scattering processes. This is consistent with the present observation of a dominant Nyquist contribution for T$<$20K. Thus, since $\tau_\varphi$ is determined by the inelastic scattering lifetime $\tau\ped{inel}$\cite{BeenakkerWL}, we can write the T dependence of $\tau_\varphi$ in the entire 4-90K range as:
\begin{equation}
\tau_\varphi^{-1}=\tau\ped{inel}^{-1}|\ped{e-e}=A\cdot \mathrm{T}+B\cdot \mathrm{T}^2
\label{full_tauphi}
\end{equation}
By inserting Eq.\ref{full_tauphi} in the first terms of the Drude conductance of Eq.\ref{full_cond}, we can fit the curves at different doping in the intermediate T region (30-90K) where the WL contribution can be neglected. With the previously determined values of $P(n\ped{2d})$ and $\tau\ped{tr}(n\ped{2d})$, this fit provides us the parameters $A(n\ped{2d})$ and $B(n\ped{2d})$. Finally, combining $\tau_\varphi(\mathrm{T})$ determined here with $\tau_\varphi(\mathrm{T})/\tau\ped{i}$ obtained from the low T (4-20K) region, we can extract $\tau\ped{i}(n\ped{2d})$. Fig.\ref{fig:5-1}b shows the excellent agreement between the fitting model (black dashed lines) and the experimental data in the WL region (solid lines).

We first apply this procedure to the R$\ped{S}$ as a function of T curves of the device on which the magnetoresistance was measured, obtaining $\tau_\varphi=39.7$ps and  $\tau_i=49.7$ps at 4K, in agreement with the values coming from the magnetoresistance fit at 2K. Once again, the comparison is carried out at 4 instead of 2K due to the signal saturation for T$\lesssim$4K in Fig.\ref{fig:4}b. This analysis is then repeated on the 4LG and 5LG devices for different values of the applied gate voltage.
\subsection{Scattering lifetimes vs. Carrier density}\label{sect:tau_doping}
We now discuss the measured trends of $\tau_\varphi$, $\tau\ped{i}$ and $\tau\ped{tr}$ as a function of $n\ped{2d}$.

In standard 2DEGs, such as GaAs heterostructures\cite{Hansen}, $\tau\ped{tr}$ is equivalent to the elastic scattering lifetime $(\tau\ped{tr}=\tau\ped{e})$ and is found to follow a dependence on $n\ped{2d}$ of the type $n^\gamma, 1 < \gamma < 2$\cite{Hansen,PaalanenPRB1984,WalukiewiczPRB1984}. This implies that $\tau\ped{tr}$ increases for increasing $n\ped{2d}$ due to the increased screening of elastic scatterers by the increased density of charge carriers\cite{Giuliani05,Mahan00}. In our devices, however, $\tau\ped{tr}$ does not follow this behavior, suggesting a significant increase of the number of elastic scatterers with increasing gate voltage.

\begin{figure}
\centerline{\includegraphics[width=80mm]{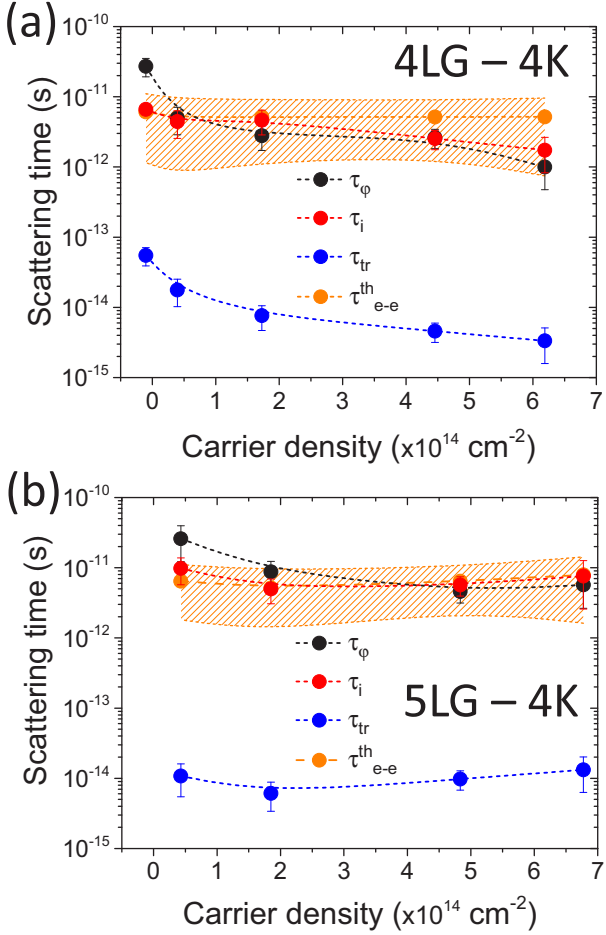}}
\caption{a) Carrier density dependence of $\tau_\varphi$ (black dots), $\tau\ped{i}$ (red dots), $\tau\ped{tr}$ (blue dots) and $\tau\ped{\varphi}|\ped{e-e}\apex{th}$ (orange dots and shaded band) at 4K in 4LG; d) The same as in a) but for 5LG. Dashed lines act as guides to the eye.}
\label{fig:5-2}
\end{figure}

If we consider the 4LG device first, we observe that $\tau\ped{tr}$ shows a monotonically decreasing behavior (Fig.\ref{fig:5-2}a, blue dots). One possible explanation was suggested in Ref.\cite{DasSarma14}, where the authors reported a decreasing mobility at high $n\ped{2d}$ for 2DEGs: In contrast to Coulomb disorder, short-range disorder would become stronger with increasing $n\ped{2d}$, thus increasing the scattering rate\cite{DasSarma14}. However, we think that this is not the main source of the increase of $\tau\ped{tr}$ in our case. A degradation of the mobility of EDL-gated devices was reported in strontium titanate\cite{GallagherNComms2015} and rhenium disulfide\cite{OvchinnikovNComms2016}, where it was suggested that ions in the electrolyte act as charged impurities. Thus, we attribute this effect to the microscopic dynamics of how EDL gating induces extra carriers in the material. EDL gating is able to induce modulations in excess of 10$^{14}$cm$^{-2}$ to the charge carrier density of a material by accumulating a densely packed layer of ions in close proximity ($\sim$1nm\cite{UenoReview2014,FujimotoReview2013}) to its surface. Refs.\cite{Zhang09,StrasserNanoLett2015} reported that charge-donating impurities at the SLG surface act as scattering centers.

We thus propose that the Li$\apex{+}$ ions in the EDL perform a similar role. Their increased density with V$\ped{G}$ competes with the increased screening induced by the extra carriers and this determines $\tau\ped{tr}$. In 4LG, the former effect would be stronger in the entire $n\ped{2d}$ range, resulting in the monotonic reduction of $\tau\ped{tr}$. If this is the case, we expect 5LG to be less sensitive to the extra scattering centers. These are localized at the surface of the first layer, and Ref.\cite{GallagherNComms2015} showed that introducing a thin spacer between the ions and the conductive channel greatly improves carrier mobility. With respect to 4LG, 5LG features one further conductive channel due to the fifth SLG, which is thus further separated from the scattering centers. Indeed, this is what we observe: The 5LG device shows an initial decrease for $n\ped{2d} \lesssim 2\cdot 10^{14}$cm$^{-2}$ followed by an increase for $n\ped{2d} \gtrsim 4\cdot 10^{14}$cm$^{-2}$ (Fig.\ref{fig:5-2}b, blue dots). Overall, the $\tau\ped{tr}$ in 5LG is nearly constant with the carrier density, as if the two competing effects on the carrier mobility were almost canceling out.

Let us now consider the doping dependence of $\tau_\varphi$ and $\tau\ped{i}$ in 4LG and 5LG, in order to allow for a direct comparison with the corresponding values for 3LG estimated from magnetoresistance. Fig.\ref{fig:5-2} shows that the doping dependence of $\tau\ped{i}$ is weakly decreasing or nearly constant within the uncertainty range: from $6.6 \pm 0.7$ps to $1.7 \pm 0.9$ps for 4LG, from $9.8 \pm 4.1$ps to $7.6 \pm 5.0$ps for 5LG. This dependence was already observed\cite{Tikhonenko09,Baker_lphi}, albeit for $n\ped{2d} \lesssim 1.5\cdot 10^{13}$cm$^{-2}$, much lower than the carrier densities considered here ($n\ped{2d} \gtrsim 1\cdot 10^{13}$cm$^{-2}$). In contrast, the doping dependence of $\tau_\varphi$ is more significant, as it is monotonically decreasing in the entire range in 4LG, while it shows a small inversion in slope on the final point for 5LG. These results are in contrast with earlier findings on FLG for $n\ped{2d} < 2\cdot 10^{13}$cm$^{-2}$\cite{Baker_lphi}. As already pointed out, e-e Nyquist scattering is the dominant dephasing mechanism in the low T range. Therefore, we can write\cite{Hansen}:
\begin{equation}
\tau\ped{\varphi}^{-1}|\ped{e-e}(\mathrm{T}) = \frac{k\ped{B} \mathrm{T}}{2 \hbar} \frac{\ln \left(\mathrm{x}\right)}{\mathrm{x}} ,\qquad \mathrm{x}=\frac{\mathrm{E}\ped{F}\tau\ped{e}}{\hbar}.
\label{tau_ee_N}
\end{equation}
The condition $\mathrm{x} \sim 1$ (Ioffe-Regel criterion\cite{IoffeRegel}) characterizes the Anderson metal-to-insulator transition\cite{MottDavis}. The metallic regime corresponds to $\mathrm{x} \gg 1$, leading to a $\tau_\varphi$ increasing with $n\ped{2d}$ (i.e. with increasing E$\ped{F}$). Under strong localization conditions, x$<$1 and the Boltzmann model no longer holds\cite{AndersonPR1958}. Intermediate x values mark a crossover from strong to weak localization\cite{MottBook} and correspond to a region where $\tau_\varphi$ can present a decreasing behavior with $n\ped{2d}$. The exact form of the $\tau_\varphi(n\ped{2d})$ curves is determined by the dependence of E$\ped{F}$ with $n\ped{2d}$ (which, in the linear regions of the bands, is a square root\cite{CastroNetoReview}) and by $\tau\ped{e}$. If we compute x corresponding to our data by using the $\tau_\varphi$ values determined at 4K, we find that for 4LG x=2.7$\pm$0.2 in the entire range of negative charge induction, while for 5LG x$<$2.7 for V$\ped{G}$=1 and 2V, x$>$5 for V$\ped{G}=3$ and 4V. Thus, the 4LG device is always in the ''crossover'' condition, while in 5LG the typical 2DEG behavior is restored for $n\ped{2d} \gtrsim 2\cdot10\apex{14}$cm$\apex{-2}$. Due to the dependence of x on $\tau\ped{e}$, these results are consistent with the fact that the ion dynamics in the EDL gating provides a competition between an increase of the number of elastic scattering center and of $n\ped{2d}$ at the increase of V$\ped{G}$. In 4LG the concurrent increase of E$\ped{F}$ and decrease of $\tau\ped{tr} = 2\tau_e$ (see Fig.\ref{fig:5-2}a) could maintain x almost constant. In 5LG at $n\ped{2d} \gtrsim 2\cdot10\apex{14}$cm$\apex{-2}$ the increase of both E$\ped{F}$ and $\tau\ped{tr}$ (see Fig.\ref{fig:5-2}b) would eventually lead to the increase of x. This particular behavior could be due to the differences in both surface-to-volume ratio and bandstructure of 4LG and 5LG, and is in agreement with the behavior of $\tau\ped{tr}$ in the two samples.

In order to further analyze these results, we evaluate the theoretical values of the e-e scattering lifetimes by using the definition of x given in Eq.\ref{tau_ee_N} with $\mathrm{E}\ped{F}(n\ped{2d})$ determined by DFT and $\tau_e$ taken from the determined $\tau\ped{tr}$ values. The results are shown in Figs.\ref{fig:5-2} as orange circles and shaded bands (that represent the uncertainty regions). Our $\tau_\varphi$ (black dots) are in agreement with the theoretical predictions within an order-of-magnitude. However, the latter are nearly doping-independent in contrast with experiments. This different behavior is particularly evident in 4LG. As expected, the addition of the Coulomb scattering term\cite{Tikhonenko09} does not affect appreciably the predicted values at this low T. Since the theoretical $\tau\ped{e-e}$ are dependent on the details of the \textit{ab-initio} calculations, the inconsistency might be related to the approximations employed in the DFT calculations of the effects of the high electric field. Another reason for the mismatch could be the validity of the hypotheses underlying the theoretical derivation of the Nyquist scattering. In particular, both the single-band and the parabolic-dispersion assumptions do not hold in FLG, particularly when the Fermi level is so far from the Dirac point\cite{Nakamura08,Koshino10}.

A comparison with the other experimental technique able to probe the e-e scattering lifetime, i.e. pump-probe spectroscopy, is not immediate, due to the different conditions in the probed electron system between transport and spectroscopic measurements. Low-bias ($<0.2$V\cite{LazzeriPRL2005}) transport measurements, such as the ones performed in this work, involve only scattering processes between carriers very close to the Fermi level. Pump-probe measurements, on the other hand, generate a strongly out-of-equilibrium carrier population by optical excitation. This population first thermalizes to a quasi-equilibrium distribution of "hot" carriers via e-e\cite{BridaNComms2013,BreusingPRB2011,ObraztsovNanoLett2011} and e-optical phonons\cite{LazzeriPRL2005,BridaNComms2013,LuiPRL2010,BreusingPRB2011,SunPRL2008,HuangNanoLett2010} scattering processes. The "hot" carriers then cool down to the rest of the Fermi sea over a much longer time scale by scattering with optical phonons\cite{LazzeriPRL2005,BridaNComms2013,LuiPRL2010,BreusingPRB2011,SunPRL2008,HuangNanoLett2010,ObraztsovNanoLett2011}. This makes the two types of measurements hardly comparable, as the e-e scattering lifetime is heavily dependent on the distance in energy between the scattering electrons and the Fermi level\cite{LiPRB2013,HwangPRB2007}. Moreover, pump-probe experiments are typically performed at room T\cite{BridaNComms2013,LuiPRL2010,BreusingPRB2011,HuangNanoLett2010,ObraztsovNanoLett2011}, where the Coulomb term dominates the e-e scattering, while our experiments are performed at low T, where the Nyquist term dominates instead. Even with all these caveats, we can try to extrapolate to room T the e-e scattering lifetimes in our systems using Eq.\ref{tau_ee_N} and imposing T=300K. This gives extrapolated room T $\tau\ped{e-e}$ between $\sim$1 and 7fs, in agreement with existing literature\cite{BridaNComms2013,BreusingPRB2011,ObraztsovNanoLett2011}.
\section{Conclusions}
We explored the electronic transport properties of 3-, 4- and 5-layer graphene in the doping regime in excess of$\sim10^{13}-10^{14}$cm$\apex{-2}$ and in the 4-30K temperature range. We used electric double layer gating to dope the samples up to$\sim7\cdot 10^{14}$cm$\apex{-2}$. We found evidence of quantum coherent transport in the entire carrier density range. Magnetoresistance measurements showed that, in the 4-30K range, transport is dominated by weak localization in the diffusive regime, and that the behavior of 3LG, 4LG and 5LG is described by the theoretical models developed for SLG and 2LG. We combined the experimental results with DFT calculations to determine the carrier scattering lifetimes as a function of the carrier density for different number of layers, and determined that electron-electron scattering with small momentum transfer (Nyquist process) is the main source of dephasing at low temperatures. Both the transport scattering lifetime and the phase coherence lifetime show a non-trivial dependence on n$\ped{2d}$. We explained the behavior of $\tau\ped{tr}$ in terms of a competing modulation of doping and density of charged scattering centers induced by EDL gating. The doping dependence of $\tau_\varphi$ points to a gate-tunable crossover from weak to strong localization, highlighting the limits of applying single-band models to multi-band systems, such as heavily doped few-layer graphene.
\section{Acknowledgments}
We thank M. Calandra, M. Polini and V. Brosco for useful discussions. We acknowledge funding from EU Graphene Flagship, ERC Grant Hetero2D, EPSRC Grant Nos. EP/ 509K01711X/1, EP/K017144/1, EP/N010345/1, EP/M507799/ 5101, and EP/L016087/1 and the Joint Project for the Internationalization of Research 2015 launched by Politecnico di Torino under funding of Compagnia di San Paolo.

\end{document}